\documentclass{article}
\usepackage{spconf,amsmath,graphicx}

\usepackage{epsfig,amssymb,subfigure,version,fancybox,mathrsfs,amscd}
\usepackage{cases,bm,multirow}
\usepackage{algorithm,algorithmic}
\usepackage{epstopdf}
\usepackage{url,xcolor}
\usepackage[colorlinks,urlcolor=blue]{hyperref}

\graphicspath{{images/}}

\newenvironment{sumMy}[1][]{\textstyle\sum}

\title{Iterative Joint Ptychography-Tomography with Total Variation Regularization}

\twoauthors
  {Huibin Chang \sthanks{NSFC 11871372 }}
	{Tianjin Normal University\\
	School of Mathematical Sciences\\
	Tianjin, China}
  {Pablo Enfedaque, Stefano Marchesini\sthanks{CAMERA, a joint ASCR-BES funded project within the Office of Science, DOE( DOE-DE-AC03-76SF00098)}}
	{Lawrence Berkeley National Laboratory\\
	Computational Research Division\\
	Berkeley, U.S.A.}

\graphicspath{{images/}}

\begin{document}
%
\maketitle
\begin{abstract}

In order to determine the 3D structure of a thick sample, researchers have recently combined ptychography (for high resolution) and tomography (for 3D imaging) in a single experiment. 2-step methods are usually adopted for reconstruction, where the ptychography and tomography problems are often solved independently. In this paper, we provide a novel model and ADMM-based algorithm to jointly solve the ptychography-tomography problem iteratively, also employing total variation regularization. The proposed method permits large scan stepsizes for the ptychography experiment, requiring less measurements and being more robust to noise with respect to other strategies, while achieving higher reconstruction quality results. 
\end{abstract}

\section{Introduction}
Ptychography \cite{nellist1995resolution} is an increasingly popular imaging technique. It overcomes the resolution limit imposed by image-forming optical elements of a regular microscope, enabling high resolution images  with a large field of view and increased contrast.  
In the recent years, ptychography has been combined with tomography~\cite{natterer1986mathematics} to produce 3D reconstructions with phase contrast and an increased resolution, providing chemical, magnetic or atomic information about the measured sample. 

In a Ptychography-Tomography (PT) experiment, a collection of diffraction patterns is measured at each rotation angle. The experiment permits to separately solve each 2D ptychography problem first, and then reconstruct the 3D volume based on the collection of ptychographic images. We refer to this analysis as 2-step approach. A downside of this scheme is that the redundancy from ptychography measurements from different rotation angles is not exploited. A joint PT model that iteratively refines the whole 3D object at the same time could exploit this additional redundancy. Such method could achieve high reconstruction quality even with low redundancy, while providing enhanced stability and robustness to noisy measurement. 

 High resolution 3D reconstructions at nano-scales were obtained in \cite{dierolf2010ptychographic,yu2018three}, employing a 2-step algorithm to solve the PT problem \cite{guizar2011phase,holler2014x,venkat2016,odstrcil2018high,Sala:19}. A recent article proposed an iterative algorithm for PT while treating each reconstruction model independently~\cite{gursoy2017direct}. In our previous work \cite{chen2017data}, an iterative algorithm was designed for near-field PT using a sandpaper analyzer in the Fresnel region. 
 
 In this paper, we propose a novel joint model that combines both ptychography and tomography problems in a standard far-field setting. We design a fast iterative algorithm based on Alternating Direction Method of Multipliers (ADMM), where each subproblem can be efficiently solved either by the Conjugate Gradient (CG) method or by simple algebraic operations.
 We employ a sparse prior in order to reduce redundancy requirements, especially when considering large stepsizes for each ptychography scanning. In addition, we consider more efficient objective functions based on the maximum likelihood  estimate of the noise, which is more accurate and produces enhanced reconstruction results. Experiments presented here demonstrate the computational performance and quality of the method:  a synthetic Shepp-Logan sample of size 128$^3$ can be reconstructed obtaining sharp features and clean background employing only 12 rotation angles and a ptychography stepsize twice as big as the illumination size.

This paper is structured as follows. In Section 2, the forward joint PT model is presented, along with the proposed ADMM-based algorithm for PT (APT). Section 3 presents the experimental results and Section 4 concludes this work.  

\section{Proposed method}
\subsection{Forward model}

We first introduce the model for 2D ptychography.  Mathematically, a 2D detector measures a collection of size $J$ of phaseless signals  $\{I_j\}_{j=0}^{J-1}$ from the far-field propagation of an illumination (probe)\;  $\omega\in \mathbb C^{\bar m}$ going through a sample $v\in \mathbb C^n$. This propagation can be modeled as 
$
I_j=|\mathcal A_j(\omega,v)|^2,
$
where  bilinear operator 
 $\mathcal A_j$ is denoted as 
 $\mathcal A_j(\omega,v):=\mathcal F(\omega\circ \mathcal S_j v),$ $\mathcal F$ represents the normalized discrete Fourier transform, $\mathcal S_j$ denotes  a binary matrix that extracts a small
window with index $j$ and size $\bar m$ over the entire image $v$,  and $\circ$ is a pointwise multiplication.

For a thick  3D sample $u$, following the Rytov approximation when the  size of the probe is much larger than the X-ray wavelength, the probe $\omega$
illuminates the linear projection of the sample $u$ rotated  at an angle $\theta$, such that the phaseless measurements $\{f_{j, \theta}\}_{j, \theta}$ are:
\begin{equation}
f_{j,\theta}= |\mathcal A_j(\omega, \mathcal Q \mathcal R_\theta u)|^2,
\label{PtychoPR}
\end{equation}
where
$\mathcal Q$ denotes the X-ray transform linear projection and $\mathcal R_\theta$ denotes the rotation transform at the angle $\theta$.


\subsection{ADMM-based algorithm for PT}
\subsubsection{Optimization Model}
Let $
\mathcal D_{j,\theta}(\omega,u):= \mathcal A_j(\omega, \mathcal Q \mathcal R_\theta u)
$.
Instead of directly solving the quadratic multidimensional systems in \eqref{PtychoPR}, following \cite{chang2019blind},
a nonlinear least squares model for PT can be addressed as
$\min\limits_{u}\tfrac12\sum\nolimits_{j, \theta}\big\||\mathcal D_{j, \theta}(\omega,u)|-\sqrt{f_{j, \theta}}\big\|^2.$
In order to deal with data contaminated by different noise types,  based on the maximum likelihood estimation (MLE), a more general mapping $\mathcal B(\cdot, \cdot): \mathbb R^{m}_+\times \mathbb R^{m}_+\rightarrow \mathbb R_+$ was introduced in \cite{chang2019blind}. It measures the distance between the recovered intensity $g\in\mathbb R_+^m$  and the collected intensity $f\in \mathbb R_+^m$  as
\begin{equation}\label{DF}
\!\mathcal B(g,f)\!=\!
\left\{
\begin{split}
& \tfrac{1}{2}\| \sqrt{g+\varepsilon \bm 1}\!-\!\sqrt{f+\varepsilon \bm 1} \|^2, \mbox{~(pAGM)}\\
&\tfrac12\langle g+\varepsilon \bm 1\!-\!(f+\varepsilon \bm 1)\circ \log(g+\varepsilon \bm 1), \bm 1_{m}\rangle, \mbox{~(pIPM)}\\
\end{split}
\right.
\end{equation}
where  $\sqrt{\cdot},\tfrac{~\cdot~}{~\cdot~}$ denote the element-wise square root and division of a vector, respectively,
$\mathbf 1$ represents a vector  whose elements are all  ones, and $\langle\cdot, \cdot\rangle$ denotes the $L^2$ inner product in Euclidean space. 
Note that the penalization parameter $\varepsilon$ was introduced  \cite{chang2019blind} to guarantee the Lipschitz differentiability of the objective function in order to prove the convergence.

Based on the the above mapping $\mathcal B(\cdot,\cdot)$, a general  optimization model with Total Variation (TV) regularization~\cite{rudin1992nonlinear} can be given as follows:
\begin{equation}\label{eqModel}
\min\limits_{u}\sumMy\nolimits_{j,  \theta}\mathcal G_{j, \theta}(\mathcal D_{j, \theta}(\omega,u))+\lambda\mathrm{TV}(u) ,
\end{equation}
where $\mathcal G_{j, \theta}(z_{j, \theta}):=\mathcal B(|z_{j, \theta}|^2,f_{j,\theta})$, and total variation $\mathrm{TV }(u):=\|\sqrt{|\nabla_x u|^2+|\nabla_y u|^2+|\nabla_z u|^2}\|_1,$
with $\nabla_x, \nabla_y$ and $\nabla_z$ denoting the finite difference in $x, y$ and $z$ directions for the 3D sample $u$.
To solve this nonconvex and nonsmooth  problem, an efficient ADMM \cite{glowinski1989augmented} algorithm will be adopted.

\subsubsection{ADMM-based PT (APT)}

By introducing $p=\nabla u:=(\nabla_x^T u, \nabla^T_y u, \nabla^T_z u)^T$ and $z_{j, \theta}=\mathcal D_{j, \theta}(\omega,u)$, \eqref{eqModel} can be reformulated, and consequently we derive the equivalent saddle point problem  as
\[
\begin{split}
&\max\nolimits_{\Lambda_1, \{(\Lambda_2)_{j,\theta}\}}\min\nolimits_{u,\{z_{j, \theta}\},p;\Lambda_1,\{{(\Lambda_2)}_{j, \theta}\}} \mathscr L(u,z,p;\Lambda_1,\Lambda_2)\\
\end{split}
\]
with augmented Lagrangian
\[
\begin{split}
& \mathscr L(u,z,p;\Lambda_1,\{(\Lambda_2)_{j,\theta}\}):=\sumMy\nolimits_{j, \theta}\mathcal G_{j,\theta}(z_{j, \theta})+\lambda \|p\|_1\\
&+r_1\Re(\langle p-\nabla u, \Lambda_1\rangle)+\tfrac{r_1}{2}\|p-\nabla u\|^2\\
\!&\!\!+\!\sumMy\nolimits_{j, \theta} \Big(r_2\Re(\langle z_{j, \theta}\!-\!\mathcal D_{j, \theta}(\omega,u), (\Lambda_2)_{j, \theta} \rangle)\!+\!\tfrac{r_2}{2}\|z_{j, \theta}\!-\!\mathcal D_{j,\theta}(\omega,u)\|^2\Big),
\end{split}
\]
and
 multipliers $\Lambda_1, \{(\Lambda_2)_{j, \theta}\}$.
The ADMM scheme is constructed by alternating the minimization of the variables $u, \{z_{j,\theta}\}, p$ and the multipliers update of $\Lambda_1, \{(\Lambda_2)_{j, \theta}\}$. 
Then, the subproblem w.r.t.  $u$ is given as
\[
\min\nolimits_u \tfrac{r_1}{2}\|\Lambda_1+p-\nabla u\|^2+\sumMy\nolimits_{j, \theta}\tfrac{r_2}{2}\|(\Lambda_2)_{j, \theta}+z_{j,\theta}-\mathcal D_{j, \theta}(\omega,u)\|^2.
\]
Then, we have
\begin{equation}\label{eqU}
\begin{split}
0&= r_1(-\Delta u+\nabla\cdot(\Lambda_1+p))\\
&\quad+r_2 \sumMy\nolimits_{j, \theta}(\nabla_u \mathcal D_{j,\theta})(\mathcal D_{j, \theta}(\omega,u)-(\Lambda_2+z)_{j, \theta}),
\end{split}
\end{equation}
with $\nabla\cdot$ denoting the divergence operator.
Introducing  operators  $\mathcal T_{j,\theta}$ as $\mathcal T_{j,\theta}u:=
\mathcal S_j \mathcal Q {\mathcal R}_{\theta} u$, 
we have
$
\mathcal D_{j,\theta}(\omega,u)=
\mathcal F ( \omega \circ \mathcal T_{j,\theta} u)$. 
Readily, \eqref{eqU} reduces to
\begin{equation}\label{eqU-3}
\begin{split}
\!\!\!&\!-r_1\Delta u+r_2 \sumMy\nolimits_{j,\theta}\mathcal T_{j,\theta}^T(|\omega|^2\circ \mathcal T_{j,\theta} u)
=-r_1\nabla\cdot(\Lambda_1\\
&\hskip 1cm+p)+r_2\sumMy\nolimits_{j,\theta}\mathcal T_{j,\theta}^T(\omega^*\circ \mathcal F^*(\Lambda_2+z)).
\end{split}
\end{equation}
We rewrite the linear systems as
\[\mathcal L u=-r_1\nabla\cdot(\Lambda_1+p)+r_2\sumMy\nolimits_{j,\theta}\mathcal T_{j,\theta}^T(\omega^*\circ \mathcal F^*(\Lambda_2+z)),\]
where the linear operator $\mathcal L$ is denoted by 
\begin{equation}
\label{eqLST}
\mathcal L u:=
-r_1\Delta u+r_2 \sumMy\nolimits_\theta(\mathcal Q \mathcal R_\theta)^T\sumMy\nolimits_j \mathcal S_j^T |\omega|^2\circ(\mathcal Q \mathcal R_\theta u).
\end{equation}
One readily knows that the above linear operator $\mathcal L$ is symmetric  positive definite such that the conjugate gradient method can be exploited.


The subproblem w.r.t. the variable $z_{j, \theta}$ is formulated as:
\[
z_{j, \theta}^\star:=\arg\min\nolimits_{z_{j, \theta}}\mathcal G_{j, \theta}(z_{j, \theta})+\tfrac{r_2}{2}\|(z+\Lambda_2)_{j, \theta}-\mathcal D_{j, \theta}(\omega,u)\|^2.
\]
Here we focus on the case where the penalization parameter $\varepsilon=0$, since the related subproblem has a closed-form solution. Readily one gets 
$
z_{j,\theta}=\mathrm{Prox}_{\mathcal G; r_2}(\tilde z_{j, \theta})
$~\cite{chang2018variational},
where 
\[
\!\!\!
\mathrm{Prox}_{\mathcal G; r_2}(\tilde z_{j, \theta})\!\!:=\!\!
\left\{
\begin{split}
&\!\tfrac{\sqrt{f_{j,\theta}}+r_2 |\tilde z_{j,\theta}|}{1+r_2}\circ \mathrm{sign}(\tilde z_{j,\theta}) \\
&\!\tfrac{r_2|\tilde z_{j,\theta}|+\sqrt{r_2^2|\tilde z_{j,\theta}|^2+4(1+r_2)f_{j,\theta}}}{2(1+r_2)}\circ \mathrm{sign}(\tilde z_{j,\theta})
\end{split}
\right.
\]
for different metrics in \eqref{DF}, 
with $\tilde z_{j,\theta}:=\mathcal D_{j, \theta}(\omega, u)-(\Lambda_2)_{j, \theta}.$ Here $\mathrm{sign}$ denotes the pointwise complex sign  of a vector.


The subproblem w.r.t. the variable $p$ is formulated as
\[
p^\star:=\arg\min_p \tfrac{\lambda}{r_1}\|p\|_1+\tfrac{1}{2}\|p+\Lambda_1-\nabla u\|^2.
\]
The soft thresholding solution is given as 
$
p^\star=\max\{0, |\nabla u-\Lambda_1|-\tfrac{\lambda}{r_1}\}\circ \mathrm{sign}(\nabla u-\Lambda_1).
$

Algorithm 1 summarizes the method and subproblems introduced above, with penalization parameter $\varepsilon=0.$ 
\begin{center}
\hrule
\vskip .03in
\centering  \textbf{Algorithm 1: APT} 
\vskip .03in
\hrule
\begin{itemize}
\item Step 0:
Initialization of $u^0=\bm 1$, $z^0_{j,\theta}=\mathcal D_{j,\theta}(\omega, u^0)$, $p=0,$  $\Lambda_1^0=0, (\Lambda_2^0)_{j,\theta}=0,$ $k=0, 
W:=\sumMy\nolimits_j \mathcal S_j^T |\omega|^2.$

\item Step 1:
Solve $u^{k+1}$ by using CG for
\vskip -.15in
\[
\begin{split}
&\!\!-r_1\Delta u+r_2 \sumMy\nolimits_\theta\mathcal P_\theta^T(W\circ\mathcal P_\theta u)=-r_1\nabla\cdot(p^k+\Lambda^k_1)\\
&\quad\qquad\quad+r_2\sumMy\nolimits_{j,\theta} \mathcal P_\theta^T\mathcal S_j^T(\omega^*\circ \mathcal F^*(z^k+\Lambda^k_2)_{j,\theta}),
\end{split}
\]\vskip -.1in
with $\mathcal P_{\theta}:=\mathcal Q\mathcal R_\theta$.
\item Step 2:
Solve $z^{k+1}_{j, \theta}$ and  $p^{k+1}$ in parallel by
\vskip -.15in
\[
\begin{split}
&z^{k+1}_{j,\theta}=\mathrm{Prox}_{\mathcal G, r_2}(\tilde  z^k_{j,\theta});\\
&p^{k+1}=\max\{0, |\tilde p^k|-\tfrac{\lambda}{r_1}\}\circ \mathrm{sign}(\tilde p^k);
\end{split}
\]\vskip -.1in
with $\tilde z^k_{j, \theta}\!:=\!\mathcal D_{j, \theta}(\omega,u^{k+1})\!-\!(\Lambda^k_2)_{j, \theta},$  $\tilde p^k:=\nabla u^{k+1}\!-\!\Lambda^k_1$.

\item Step 3:
Update the multipliers by
\vskip -.15in
\[
\begin{split}
&\Lambda^{k+1}_1= \Lambda^k_1+p^{k+1}-\nabla u^{k+1};\\
&(\Lambda^{k+1}_2)_{j, \theta}= (\Lambda^k_2)_{j, \theta}+z^{k+1}_{j, \theta}-\mathcal D_{j, \theta}(\omega,u^{k+1}).
\end{split}
\]
\vskip -.3in
\item Step 4:
If satisfying the convergence condition, stop and output $u^{k+1}$ as the final result; otherwise, go to Step 1.
\vskip -.4in
\end{itemize}
\hrule
\end{center}


We also consider a special case without regularization with $\lambda=0$ in \eqref{eqModel}. Then the APT algorithm can be further simplified by setting $\lambda=r_1=0$ in Step 1 of Algorithm 1, and removing the update of variables $p$ and $\Lambda_1$. This special case is denoted as APTs (APT simplified, without TV).


The current algorithm can be  generalized to a problem with unknown probe. Similarly to \cite{chang2019blind}, one has to solve the subproblem w.r.t. the probe, i.e.
$\min_\omega\sum\nolimits_{j,\theta} \tfrac{1}{2}\|z_{j,\theta}+(\Lambda_2)_{j,\theta}-\mathcal D_{j,\theta}(\omega,u)\|^2.
$
Calculating the first-order derivative yields
$\omega\leftarrow\tfrac{\sum_{j,\theta}\left(\mathcal S_j \mathcal Q \mathcal R_\theta u\right)^*\circ\mathcal {\mathcal F}^{*}(z+\Lambda_2)_{j,\theta} }{\sum_{j,\theta} |\mathcal S_j \mathcal Q\mathcal R_\theta  u|^2}.$

We have presented the algorithm for the special case where $\varepsilon=0$ in \eqref{DF}. The Lipschitz differentiability of the objective functional can be used to show the convergence of the APT algorithm, similarly to \cite{chang2019blind}.
The Lipschtiz property of the objective requires a positive penalization parameter $\varepsilon>0$. This case requires an extra inner loop, leading to additional computational cost. Numerically we observe convergence of APT using $\varepsilon=0$, and we leave the convergence analysis for APT without Lipschtiz property as future work. 

\begin{figure}[h]
\vskip -.1in
\begin{center}
\subfigure[]{\includegraphics[width=.1\textwidth]{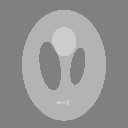}}
\qquad\qquad
\subfigure[]{\includegraphics[width=.05\textwidth]{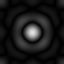}}
\end{center}\vskip -23pt
\caption{(a) Ground truth synthetic image from a 3D Shepp-Logan phantom, slice 64 out of 128. (b) Absolute value of the illumination (probe) with beamsize (Full width at half maximum (FWHM)) 14 pixels.}
\vskip -.2in
\label{fig0}
\end{figure}

\section{Numerical experiments}
For this experimental analysis we consider a synthetic 3D sample with pure phase with size $128\times 128\times 128$ pixels (Fig. \ref{fig0}(a)). The ptychography measurements are simulated using a single grid scan with non-periodical boundary at every rotation angle, with a  probe of size $64\times 64$ pixels (amplitude shown in Fig.\ref{fig0}(b)). For the tomography setting, the sample is rotated discretely with angles ranging from 0 to $\pi$.
Two quantitative measurements will be used in this analysis: the signal-to-noise ratio (SNR) and the R-factor. The ground truth sample can be used to compute the SNR of $u^k$ as:
 $
 \mathrm{SNR}(u^k, u_g)=-10\log_{10}\tfrac{\sum\limits_{t}|\zeta^* u^k(t+T^*)-u_g(t)|^2}{\|\zeta^* u^k\|^2},
 $
where $u_g$ corresponds to the ground truth sample. The error is computed up to the translation $T^*$ and phase factor $\zeta^*$, and the phase shift and scaling factor $\zeta^*$ are determined by
$(\zeta^*,T^*):=\arg\min\limits_{\zeta\in \mathbb C, T\in \mathbb Z}\sum_{t}|\zeta u^k(t+T)-u_g(t)|^2$.
The R-factor (fitting error)
is defined
as:
$
\text{R-factor}^k:= \tfrac{\sum\nolimits_{j,\theta}\big\||\mathcal D_{j,\theta}(\omega, u^k)|-\sqrt{f_{j,\theta}} \big\|_1}{\sum\nolimits_{j,\theta}\|\sqrt{f_{j,\theta}}\|_1}.
$

First, we show the advantage of the proposed algorithm compared with the 2-step method. The 2-step strategy is computed using SHARP~\cite{marchesini2016sharp} for the ptychography reconstruction  (with default parameters and 100 iterations) and conjugate gradient with 10 iterations for the tomography part  using the Tomopy framework \cite{tomopy}. We show the results with scan stepsize 32 pixels ($230\%$  of beamsize).
The reconstruction results for the center slice of the sample can be found in Fig. \ref{fig1}. With this configuration, the 2-step approach does not converge to a reconstruction (Figs. \ref{fig1} (a),(d)), since the ptychography solver does not have enough redundancy giving the large scan stepsizes. On the other hand, the proposed APTs algorithm (Figs. \ref{fig1} (b),(e)) is able to recover the sample but presenting some blurred areas, whereas APT can achieve a cleaner reconstruction with very sharp features (Figs. \ref{fig1} (c), (f)).  With the help of  more rotation angles, the results are also improved, as depicted in Fig. \ref{fig1} second row.  The R-factors and SNRs of the recovery results can also be found in Table \ref{tab0}, showing how APT with regularization achieves better R-factors and SNRs
than APTs.

For the next experiment, we generate a second dataset with much smaller scan stepsize for ptychography (4 pixels), and maintaining 12 rotation angles for tomography. Reconstruction results for this experiment are depicted in Fig. \ref{fig2}. We can see how the 2-step method and APTs produce comparable visual results in this case, whereas APT produces a much more nitid reconstruction with sharper features. The R-factors and SNRs for this experiment can also be found in the last column of Table \ref{tab0}, showing the improvements of the proposed  algorithms, compared with the 2-step method.

\def\figwidth{0.25\columnwidth}

\begin{figure}[]
\vskip -.3in
\begin{center}
\subfigure[]{\includegraphics[width=\figwidth]{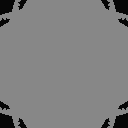}}
\subfigure[]{\includegraphics[width=\figwidth]{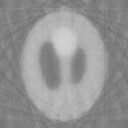}}
\subfigure[]{\includegraphics[width=\figwidth]{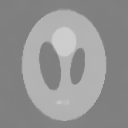}}\\
\vskip -10pt
\subfigure[]{\includegraphics[width=\figwidth]{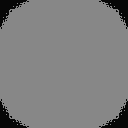}}
\subfigure[]{\includegraphics[width=\figwidth]{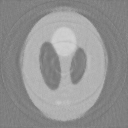}}
\subfigure[]{\includegraphics[width=\figwidth]{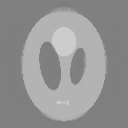}}
\end{center}\vskip -23pt
\caption{
Results of a synthetic sample generated with stepsize~=~32 and 12 rotation angles (first row) and with 48 rotation angles (second row). First column: reconstruction using the 2-step method. Second column: APTs. Third column: APT.}
\vskip -.1in
\label{fig1}
\end{figure}

\begin{figure}[h]
\vskip -.2in
\begin{center}
\subfigure[]{\includegraphics[width=\figwidth]{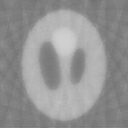}}
\subfigure[]{\includegraphics[width=\figwidth]{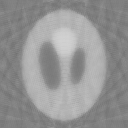}}
\subfigure[]{\includegraphics[width=\figwidth]{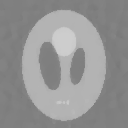}}\\
\end{center}\vskip -23pt
\caption{Reconstruction results of a synthetic sample generated with stepsize = 4 and 12 rotation angles, using (a) the 2-step method, (b) APTs and (c) APT.}
\label{fig2}
\vskip -.1in
\end{figure}

\begin{figure}[]
\vskip -.3in
\begin{center}
\subfigure[Error ]{\includegraphics[width=.2\textwidth]{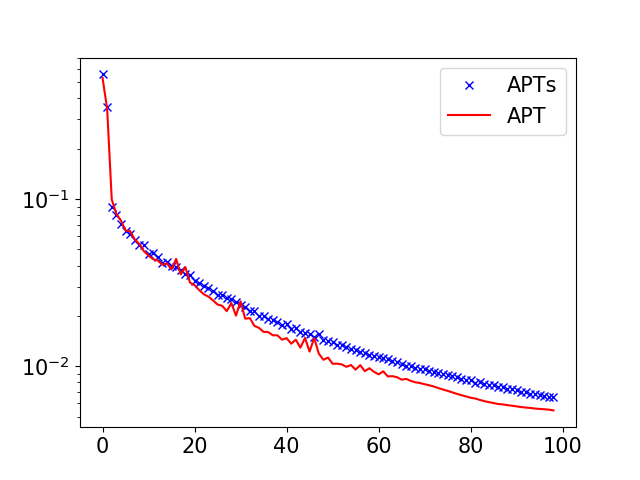}}~~~~
\subfigure[R-factor]{\includegraphics[width=.2\textwidth]{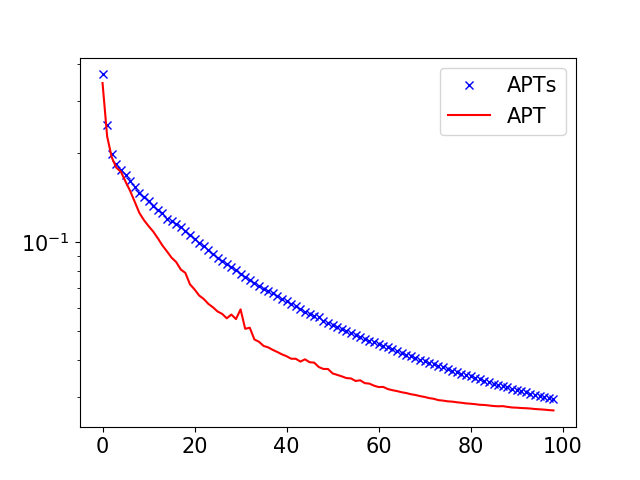}}
\end{center}\vskip -20pt
\caption{Convergence curves  with stepsize = 32 and 12 rotation angles. Horizontal axis denotes iteration count. Vertical axis: (a) Relative error as  $\tfrac{\|u^k-u^{k+1}\|}{\|u^{k+1}\|}$, (b) R-factor.} 
\label{fig3}
\vskip -.2in
\end{figure}

\begin{table}
    \caption{SNRs and R-factors for results using stepsizes = 4, 32 and rotation angles = 12, 48, using the 2-step method, APTs and APT. $*$ stands for a failed reconstruction.}
    \vskip .1in
    \centering
    \scalebox{.6}{
    \begin{tabular}{|c|c|c|c|c|}
    \hline
    \multicolumn{2}{|c|}{stepsize}&  32 & 32&4 \\
      \hline
     \multicolumn{2}{|c|}{number of angles} &  12& 48&12 \\
          \hline
         \multirow{3}{*}{ R-factor}&2-step & $\star$   &$\star$  & 0.0185 \\
                                  &APTs & 0.0295   & 0.0263 &  0.00995  \\
                                  &APT &0.0270    & 0.0189  &  0.00852 \\
         \hline
        \multirow{3}{*}{SNR}& 2-step  &$\star$    & $\star$  &16.8\\
                            &  APTs    &15.1  & 17.0  &17.6\\
                            &  APT &24.1  & 25.6  &26.9\\
         \hline
    \end{tabular}}
    \label{tab0}
    \vskip -.1in
\end{table}
The next experiment, presented in~Fig. \ref{fig3}, analyzes the convergence of the proposed algorithms. The figure shows how both the errors and R-factors steadily decrease, demonstrating the reliability and robustness of the method.

\begin{figure}[h]
\vskip -.23in
\begin{center}
\subfigure[APTs]{\includegraphics[width=.24\columnwidth]{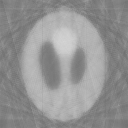}}
\subfigure[APT ]{\includegraphics[width=.24\columnwidth]{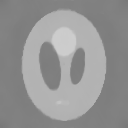}}
\subfigure[APTs]{\includegraphics[width=.24\columnwidth]{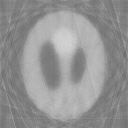}}
\subfigure[APT ]{\includegraphics[width=.24\columnwidth]{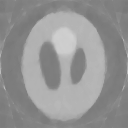}}
\end{center}\vskip -20pt
\caption{
Reconstruction results with data contaminated by Poisson noise, with stepsize = 32 and 12 rotation angles, using the proposed APTs and APT algorithms.  (a)-(b): Peak level $\eta=1$; (c)-(d): Peak level $\eta=0.1$. The quality of the noisy measurements with respect to the original intensities are $\mathrm{SNR}_{intensity}=46.3, 38.4$  for (a)-(b) and (c)-(d), respectively.  
} 
\vskip -.3in
\label{fig4}
\end{figure}

Finally, we test the robustness of the proposed algorithms when data is contaminated by noise.
In order to simulate different intensity peak values,  a factor $\eta>0$ is introduced, such that Poisson noise corrupts the clean intensity as $f_{j,\theta}(t)\stackrel{\mathrm {i.i.d.}}{\sim} \mathrm{Poisson} (f_{j, \theta}^\eta(t))$, where $f_{j, \theta}^\eta:=\eta|\mathcal D_{j,\theta}(\omega, u_g)|^2$, with ground truth $\omega$ and $u_g.$
The smaller the $\eta$ factor the bigger the noise level. $\mathrm{SNR}_{intensity}$ \cite{chang2019blind} is used to measure the noise level, which is defined as:
$
\mathrm{SNR}_{intensity}:=-10\log_{10}\tfrac{\sum\nolimits_{j,\theta}\| f_{j,\theta}-f_{j,\theta}^\eta\|^2}{\sum\nolimits_{j,\theta}\|f_{j,\theta}^\eta\|^2}.
$
We show the reconstruction results with peak levels 0.1 and 1 in Fig. \ref{fig4}, and the respective R-factors and SNRs in Table \ref{tab1}. The results demonstrate how the proposed algorithms are robust to the noisy measurements, even with significantly large scan stepsizes and few angles. As reported in the previous analysis, Fig. \ref{fig4} and Table \ref{tab1} show how APT can greatly enhance the reconstruction quality with respect to APTs thanks to the TV regularization. 

\begin{table}
\vskip -.02in
    \caption{SNRs and R-factors of the reconstruction results from data contaminated by Poisson noise, with stepsize = 32 and 12 rotation angles.}
    \centering
    \scalebox{.6}{
    \begin{tabular}{|c|c|c|c| }
      \hline
     \multicolumn{2}{|c|}{$\eta$} & 1& 0.1 \\
          \hline
         \multirow{2}{*}{ R-factor} &APTs & 0.0455   & 0.0796   \\
                                  &APT &0.0427   & 0.0732  \\
         \hline
        \multirow{2}{*}{SNR}&  APTs    & 14.9 &13.6  \\
                            &  APT &22.7  &19.4 \\
         \hline
    \end{tabular}
    }
    \label{tab1}
    \vskip -.2in
\end{table}

\section{Conclusion}

We propose a novel model and algorithm to iteratively solve the coupled PT problem for far-field X-ray imaging. We present the ADMM-based APT and APTs algorithms that are designed on subproblems requiring low computational cost, while providing higher reconstruction quality, stability and robustness. The proposed model and algorithms exploit the redundancy of ptychography measurements from multiple rotation angles, and the capabilities of the method regarding quality and convergence are demonstrated in the experimental analysis. As a future work, we will test our method using real experimental data, and also consider additional experimental issues, such as sample drifts \cite{yu2018three} and phase unwrapping~\cite{guizar2011phase}.

\bibliographystyle{IEEEbib}
\bibliography{rD}

\begin{thebibliography}{10}

\bibitem{nellist1995resolution}
PD~Nellist, BC~McCallum, and JM~Rodenburg,
\newblock ``Resolution beyond the `information limit' in transmission electron
  microscopy,''
\newblock {\em Nature}, vol. 374, no. 6523, pp. 630, 1995.

\bibitem{natterer1986mathematics}
Frank Natterer,
\newblock {\em The mathematics of computerized tomography}, vol.~32,
\newblock Siam, 1986.

\bibitem{dierolf2010ptychographic}
Martin Dierolf, Andreas Menzel, Pierre Thibault, Philipp Schneider, Cameron~M
  Kewish, Roger Wepf, Oliver Bunk, and Franz Pfeiffer,
\newblock ``Ptychographic x-ray computed tomography at the nanoscale,''
\newblock {\em Nature}, vol. 467, no. 7314, pp. 436--439, 2010.

\bibitem{yu2018three}
Young-Sang Yu, Maryam Farmand, Chunjoong Kim, Yijin Liu, Clare~P Grey, Fiona~C
  Strobridge, Tolek Tyliszczak, Rich Celestre, Peter Denes, John Joseph,
  et~al.,
\newblock ``Three-dimensional localization of nanoscale battery reactions using
  soft x-ray tomography,''
\newblock {\em Nature communications}, vol. 9, no. 1, pp. 921, 2018.

\bibitem{guizar2011phase}
Manuel Guizar-Sicairos, Ana Diaz, Mirko Holler, Miriam~S Lucas, Andreas Menzel,
  Roger~A Wepf, and Oliver Bunk,
\newblock ``Phase tomography from x-ray coherent diffractive imaging
  projections,''
\newblock {\em Optics express}, vol. 19, no. 22, pp. 21345--21357, 2011.

\bibitem{holler2014x}
M~Holler, A~Diaz, M~Guizar-Sicairos, P~Karvinen, Elina F{\"a}rm, Emma
  H{\"a}rk{\"o}nen, Mikko Ritala, A~Menzel, J~Raabe, and O~Bunk,
\newblock ``X-ray ptychographic computed tomography at 16 nm isotropic 3d
  resolution,''
\newblock {\em Scientific reports}, vol. 4, pp. 3857, 2014.

\bibitem{venkat2016}
Singanallur~V Venkatakrishnan, Maryam Farmand, Young-Sang Yu, Hasti Majidi,
  Klaus van Benthem, Stefano Marchesini, David~A Shapiro, and Alexander
  Hexemer,
\newblock ``Robust x-ray phase ptycho-tomography,''
\newblock {\em IEEE Signal Processing Letters}, vol. 23, no. 7, pp. 944--948,
  2016.

\bibitem{odstrcil2018high}
Michal Odstrcil, Mirko Holler, Jorg Raabe, and Manuel Guizar-Sicairos,
\newblock ``High resolution 3d imaging of integrated circuits by x-ray
  ptychography,''
\newblock in {\em Image Sensing Technologies: Materials, Devices, Systems, and
  Applications V}. International Society for Optics and Photonics, 2018, vol.
  10656, p. 106560U.

\bibitem{Sala:19}
Simone Sala, Darren~J. Batey, Anupama Prakash, Sharif Ahmed, Christoph Rau, and
  Pierre Thibault,
\newblock ``Ptychographic x-ray computed tomography at a high-brilliance x-ray
  source,''
\newblock {\em Opt. Express}, vol. 27, no. 2, pp. 533--542, Jan 2019.

\bibitem{gursoy2017direct}
Do{\u{g}}a G{\"u}rsoy,
\newblock ``Direct coupling of tomography and ptychography,''
\newblock {\em Optics letters}, vol. 42, no. 16, pp. 3169--3172, 2017.

\bibitem{chen2017data}
Michael Chen, Dilworth Parkinson, Singanallur Venkatakrishnan, Stefano
  Marchesini, and Laura Waller,
\newblock ``Data efficient x-ray phase tomography with self-calibrated
  sandpaper analyzer,''
\newblock in {\em Computational Optical Sensing and Imaging}. Optical Society
  of America, 2017, pp. CW2B--2.

\bibitem{chang2019blind}
Huibin Chang, Pablo Enfedaque, and Stefano Marchesini,
\newblock ``Blind ptychographic phase retrieval via convergent alternating
  direction method of multipliers,''
\newblock {\em SIAM Journal on Imaging Sciences}, vol. 12, no. 1, pp. 153--185,
  2019.

\bibitem{rudin1992nonlinear}
Leonid~I Rudin, Stanley Osher, and Emad Fatemi,
\newblock ``Nonlinear total variation based noise removal algorithms,''
\newblock {\em Physica D: Nonlinear Phenomena}, vol. 60, no. 1, pp. 259--268,
  1992.

\bibitem{glowinski1989augmented}
Roland Glowinski and Patrick Le~Tallec,
\newblock {\em Augmented Lagrangian and operator-splitting methods in nonlinear
  mechanics},
\newblock Philadelphia, PA: SIAM, 1989.

\bibitem{chang2018variational}
Huibin Chang, Stefano Marchesini, Yifei Lou, and Tieyong Zeng,
\newblock ``Variational phase retrieval with globally convergent preconditioned
  proximal algorithm,''
\newblock {\em SIAM Journal on Imaging Sciences}, vol. 11, no. 1, pp. 56--93,
  2018.

\bibitem{marchesini2016sharp}
Stefano Marchesini, Hari Krishnan, David~A Shapiro, Talita Perciano, James~A
  Sethian, Benedikt~J Daurer, and Filipe~RNC Maia,
\newblock ``{SHARP:} a distributed, {GPU}-based ptychographic solver,''
\newblock {\em Journal of Applied Crystallography}, vol. 49, no. 4, pp.
  1245锟紺--1252, 2016.

\bibitem{tomopy}
Doga G{\"u}rsoy, Francesco De~Carlo, Xianghui Xiao, and Chris Jacobsen,
\newblock ``Tomopy: a framework for the analysis of synchrotron tomographic
  data,''
\newblock {\em Journal of synchrotron radiation}, vol. 21, no. 5, pp.
  1188--1193, 2014.

\end{thebibliography}
\end{document}